\documentclass[12pt,a4paper,twoside]{article}
\usepackage{epsfig}
\usepackage{baltlat1}
\usepackage{wrapfig}
\begin{document}
\ \
\vspace{0.5mm}

\setcounter{page}{1}
\vspace{8mm}

\titlehead{Baltic Astronomy, vol.12, XXX--XXX, 2003.}

\titleb{Compact object mergers as progenitors of short bursts}

\begin{authorl}
\authorb{T.~Bulik}{1} and
\authorb{K.~Belczynski}{2,3}
\end{authorl}

\begin{addressl}
\addressb{1}{Nicolaus Copernicus Astronomical Center, Bartycka 18, 00716 Warsaw, Poland }

\addressb{2}{ Northwestern University, 2145 Sheridan Rd. Evanston, IL, USA}

\addressb{3}{Lindheimer Fellow}

\end{addressl}

\submitb{Received October 15, 2003}

\begin{abstract}
Compact object mergers are possible progenitors of short burst
We analyze  properties of compact object mergers  
  using the StarTrack population synthesis code, and find that the
double neutron star population is dominated by short lived systems, thus
they merge within host galaxies, while black hole neutron star binaries
merge outside of the host galaxies.

 \end{abstract}


\resthead{Compact object mergers as progenitors of short bursts}{Bulik \& Belczynski}



\sectionb{1}{INTRODUCTION}

Ever since the discovery (Klebesadel, Strong \& Olson 1973) of gamma-ray bursts 
(GRBs) the main problem 
faced by researchers was to identify the physical mechanism behind
these phenomena and to place them  somewhere within the realm of standard 
astrophysics. The discovery of afterglows (Costa et~al. 1997) lead to great progress
in the field. It allowed to  identify GRB host galaxies, and to study their properties. 
GRB afterglows were localized within host galaxies. Long GRBs take place
inside small star forming galaxies. 

Coalescences of compact object binaries have been considered as 
a possible mechanism of GRBs. Several types of such mergers
were considered - see Fryer, Woosley \& Hartmann (1999).
 The main problem is that the timescale of the merger is relatively short and it is difficult to explain 
long bursts, yet it is still possible that the progenitors 
of  short bursts
are   compact object mergers.

These discoveries were followed by 
several studies (Bloom, Sigurdsson \& Pols 1997; Bulik, Belczynski \& Zbijewski 1997) 
using the population synthesis 
approach to  calculate  offsets for simulated populations of possible GRB progenitors,
with conclusion that most double neutron star binaries merge outside host galaxies.

Here we report the results of the careful examination of properties of 
compact object binaries with the StarTrack population synthesis 
code (Belczynski, Kalogera \& Bulik 2002), which singificantly 
changes the conclusions reported above.

\sectionb{2}{PROPERTIES OF COMPACT OBJECT BINARIES}

A novel feature of the StarTrack code (Belczynski, Kalogera \& Bulik 2002) was to include
the evolution of He stars. Within the code helium stars with the mass under $4.5\,M_\odot$ 
develop convective envelopes. This leads to a possibility of initiating common envelope  (CE)
episodes by He stars in close binaries. 

We have found that one can identify three different types of evolutionary paths leading to 
formation of double neutron star (NSNS) binaries  (Belczynski, Bulik \& Kalogera 2002). 
Group I contains
 systems which involved a double 
CE phase between tow helium stars which later underwent supernova explosions.
Group II consist of systems which underwent a CE with a He star donor in a binary 
with a NS. Group III are the remaining, classical  systems for which   no CE phase 
with an He star took place.  The systems of Group II dominate the population 
of NSNS binaries,   they constitute nearly  $90$\% of the
entire population of such objects. However these binaries as well as the ones of Group I
are very short lived.   The additional CE phase tightens up these 
systems which decreases significantly their lifetimes as compact object binaries.
The typical lifetimes due to gravitational wave emission 
of these binaries is $\approx 1$\,Myr.  The nuclear lifetimes of their progenitors 
is about ten times longer.  On the other hand black hole neutron star
(BHNS) binaries cannot undergo 
this additional CE phase because the massive He star do not have 
convective envelopes. Therefore the BHNS binaries are much more long lived.
The NSNS binaries of Group III are also long lived, the typical lifetimes 
of BHNS and NSNS binaries of Group III stretches from a Gyr to the Hubble time. 
We present the distribution of lifetimes of BHNS and NSNS binaries 
from the formation of two stars on ZAMS to the merger due to gravitational wave emission
in Figure~1.
Since the population of NSNS binaries is dominated   by the Group II systems the 
typical lifetime of such systems stretches from $10$ to $50$\,Myrs.

\begin{figure}[t] 
\centerline{\psfig{figure=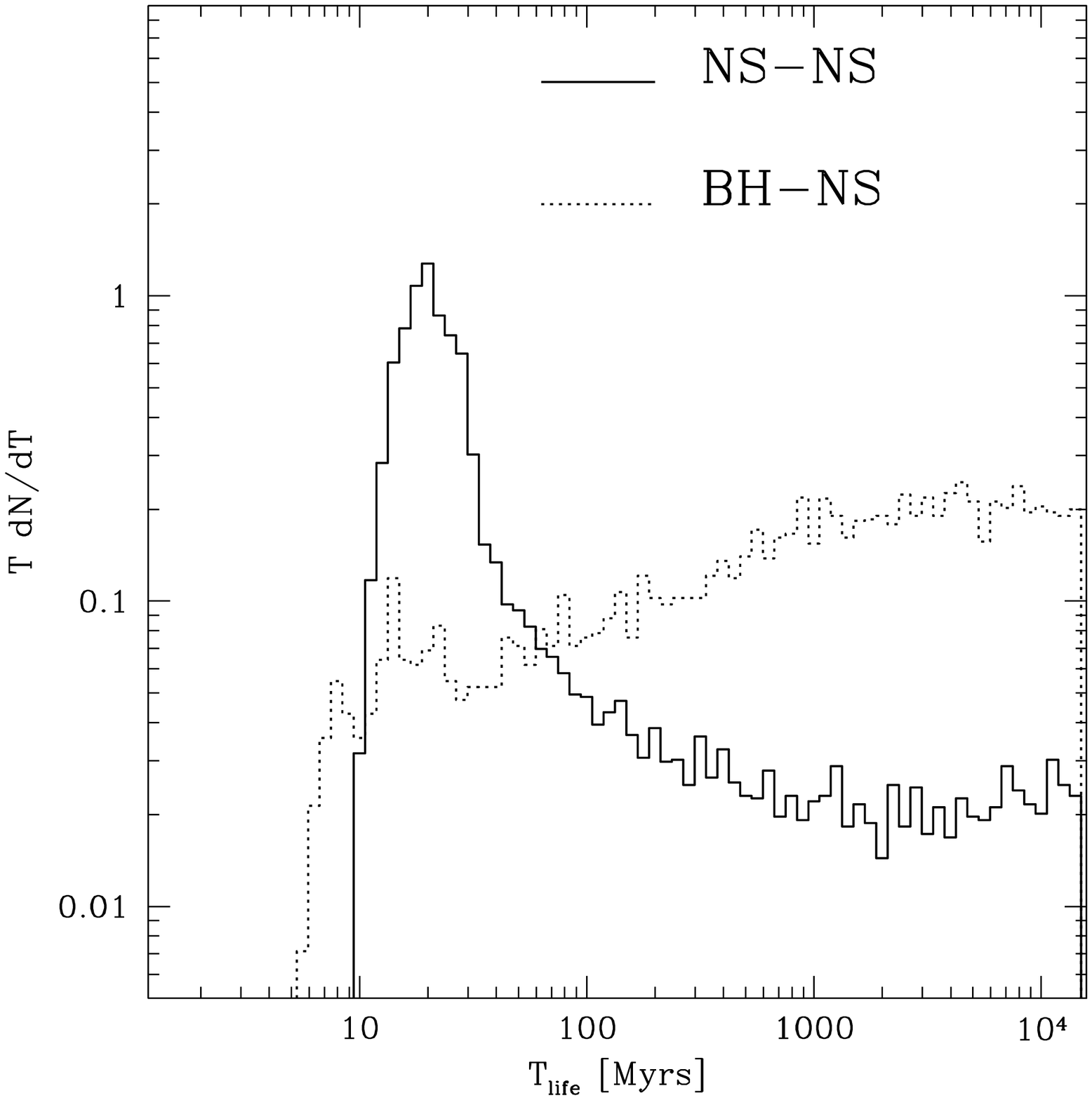,width=80truemm,angle=0,clip=}}
\captionb{1}{The distributions of the lifetimes of NS-NS binaries
ans BH-NS binaries.}
\end{figure}

Given the lifetimes and the velocities attained due to supernova
kicks by the compact object binaries one can calculate the distributions
 of their merger sites around galaxies. We have found 
that  
the NSNS binaries merger predominantly within the host galaxies,
contrary to the results of previous studies (Belczynski, Bulik \& Rudak 2002). 
This is due to the fact that 
although the spatial velocities of the NSNS binaries are of the 
order of up to few hundred km\,s$^{-1}$, their lifetimes are short
and they do not have the time to leave the host galaxies. 
The NSNS binaries of Group III manage to leave the host galaxies and 
merge outside of them.
BHNS binaries leave the host galaxies: 
 although their spatial 
velocities are smaller  their 
lifetimes are much longer.

\begin{table}
\begin{center}
\begin{tabular}{lcc}
\multicolumn{3}{c}{\parbox{11cm}{\baselineskip=8pt
~~~~{\smallbf Table 1.}{\small\ Properties of compact object binary  mergers.}}}\\
\tablerule
 ~ &   NSNS & BHNS\\
\tablerule
Merger sites          &  inside galaxies\hhuad & outside galaxies\hhuad\\[-2pt] 
 Trace star formation &   Yes\hhuad &    No \hhuad\\[-2pt]
Luminosity function  &  Narrow \hhuad& Wide \hhuad\\[-2pt]
\tablerule
\end{tabular}
\end{center}
\end{table}

\sectionb{3}{CONCLUSIONS}
 
We have  reviewed the properties of possible progenitors
of short GRBs: the coalescences of compact object binaries.
We find that the properties of the NSNS and BHNS binaries are significantly  different. 
The lifetimes of a majority of 
NSNS binaries are much  shorter then
the lifetimes of BHNS binaries. We find that the distribution of merger 
sites  of these two types of binaries are different: NSNS binaries
merge inside host galaxies, while the BHNS binaries merge outside of them. 
The difference in the lifetimes of these two types of binaries
implies also different redshift distributions of such mergers. The NSNS
binary mergers shall trace the star formation rate history, while the
BHNS mergers shall not because of their long lifetimes. 
Finally, the luminosity function, or rather the internal 
pool of energy in the burst should be much narrower for the NSNS
mergers than for the BHNS mergers. This is due to the fact that
the mass range of neutron star is rather narrow, while
the masses of black holes of stellar origin may reach 
up to $15-20\,M_\odot$ for population I stars, and could 
be much higher for the black holes originating from lower metallicity
population II and earlier stars. The   properties of NSNS and BHNS 
mergers are summarized in Table~1.

ACKNOWLEDGMENTS.\ This work was supported by the KBN grant
5P03D01120. TB thanks the organizers of the meeting for support.
 H\'al\'asan  k\" osz\" onom!   
\goodbreak

\References
\newcommand{\apj}{ApJ}
\newcommand{\mnras}{MNRAS}
\newcommand{\iaucirc}{IAU Circ.}
\newcommand{\apjl}{ApJ Lett.}

\ref Belczynski, K., Bulik, T., \& Kalogera, V.\ 2002, \apjl, 571, L147 

\ref Belczynski, K., Bulik, T., \& Rudak, B.\ 2002, \apj, 571, 394 

\ref Belczynski, K., Kalogera, V., \& Bulik, T.\ 2002, \apj, 572, 407

\ref Bloom,  J.~S., Sigurdsson, S., \& Pols, O.~R.\ 1999, \mnras, 305, 763 
\ref Bulik, T., Belczy{\' n}ski, K., \& Zbijewski, W.\ 1999, \mnras, 309, 629 

\ref Costa, E.~et al.\ 1997, \iaucirc, 6576, 1 

\ref Fryer, 
C.~L., Woosley, S.~E., \& Hartmann, D.~H.\ 1999, \apj, 526, 152 

\ref Klebesadel, R.~W., Strong, I.~B., \& Olson, R.~A.\ 1973, \apjl, 182, L85 
\end{document}